\documentstyle[twocolumn,amssymb,aps,psfig,float]{revtex}

\begin{document}
\narrowtext

\twocolumn[\hsize\textwidth\columnwidth\hsize\csname @twocolumnfalse\endcsname 

\title{On the origin of biquadratic exchange in spin 1 chains} 
\author{Frederic Mila$^1$ and Fu-Chun Zhang$^2$}

\address{$^1$ Laboratoire de Physique Quantique, Universit\'e Paul
Sabatier, 31062 Toulouse Cedex, France\\
$^2$ Department of Physics, University of Cincinnati, Cincinnati, Ohio 45221}

\maketitle

\begin{abstract}
\begin{center} 
\parbox{14cm}
{
One dimensional spin 1 systems may have a rich phase diagram including
Haldane gap and dimerized phases if the usually very small biquadratic exchange
becomes significant.  We show that this unlikely condition may be
fulfilled in electron systems with quasi-degenerate orbitals. 
This mechanism may have been experimentally realized in the spin 1 chain
LiVGe$_2$O$_6$. The implications for the exploration of the physics and quantum
chemistry of spin 1 chains are discussed.
}
\end{center}

\end{abstract}

\vskip .1truein
 \noindent PACS: 75.30.Et Exchange and superexchange interactions - 
 75.10.Jm Quantized spin models - 75.50.Ee Antiferromagnetics 
\vskip2pc
]

There has been a lot of activity recently on the physics of one-dimensional
(1D) spin 1 systems, especially after the prediction by Haldane that the 1D
Heisenberg model has a spin gap for integer spins~\cite{haldane}. That
prediction has been confirmed since then by the observation of a gap in many
spin 1 chains~\cite{overview}.  There has also been a lot of progress in
the study of the most general Hamiltonian describing an isotropic 
coupling between
neighboring spins 1, namely~\cite{affleck} 
\begin{eqnarray}
H & = & \sum_{\langle ij \rangle}H_{ij},  
\nonumber \\
H_{ij} & = & J_1 \vec S_i \cdot \vec S_{j} + J_2(\vec S_i \cdot \vec S_{j} )^2,
\label{hamiltonian}
\end{eqnarray}
where the sum $\langle ij \rangle$ is over nearest--neighbor 
pairs. The phase diagram of this model is extremely rich
~\cite{schollwock}. It is most easily described using the parametrization $J_1
=J \cos \theta$, $J_2=J \sin \theta$.  As illustrated in Fig. 1, the system has
two gapped phases: The Haldane phase for $- \pi /4 < \theta < \pi /4$ and a
dimerized phase for $-3 \pi /4 < \theta < - \pi /4$.  The two phases are
connected by a critical point at $\theta = - \pi/4$ for which the model is
exactly solvable and the spectrum is gapless. In addition,
there is a Lifshitz point and a disordered point in the Haldane phase, as well
as a Valence Bond Solid point~\cite{affleck}, and the properties for
$\pi/4<\theta<\pi/2$ and $\theta\simeq -\pi/4$ are not totally agreed upon yet.

\begin{figure}[hp]
\centerline{\psfig{figure=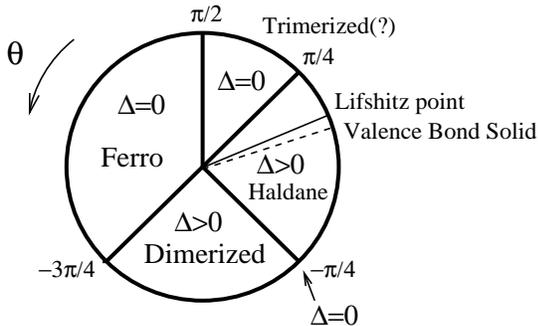,width=7.0cm,angle=0}}
\vspace{0.5cm}
\caption{Phase diagram of the general spin 1 model of Eq. (\ref{hamiltonian})
after Ref. [2].}
\label{fig1}
\end{figure}

Hopelessly, it has not been
possible until very recently to explore this phase diagram experimentally due
to the lack of systems with a sizable biquadratic exchange, which is
conveniently measured by the ratio $-\beta = J_2/J_1 = \tan \theta$. It has
been generally assumed that $\beta$ is always small so that the system is
always in the Haldane gap phase. The experiments on many spin 1 chain
compounds have indeed supported this view until a very recent experiment 
on the 1D,
spin 1 vanadium oxide LiVGe$_2$O$_6$. In that experiment, an abrupt drop in
magnetic susceptibility typical of a
spin-Peierls transition has been observed at a temperature of 22
K~\cite{millet}. This property is consistent with a gapless spectrum above this
temperature, and it has been argued that this behavior is most likely due to 
the presence of a significant biquadratic exchange interaction~\cite{millet}. 

Although biquadratic or higher-order spin exchange
interactions have been discussed in the past and shown
to induce anomalous magnetic properties in spin $S \geq 1$ systems, they are
usually small~\cite{nagaev,EuSe}.  It is therefore of interest to study the
underlying mechanism for predominant biquadratic interaction.

In this paper we propose a new microscopic mechanism leading to a significant
biquadratic interaction in spin 1 systems. We use a microscopic model to 
derive the interaction couplings $J_1$ and $J_2$ in Eq. (\ref{hamiltonian})
based on a  perturbation theory, which is compared with exact numerical
calculations.  The key in
our model is to include a third atomic orbital for an electron with slightly
higher energy than the two singly occupied lowest orbital states, i.e. to
consider the situation where we have a quasi-degeneracy of the orbitals on one
site. The virtual
electron transition via the third orbital level favors ferromagnetic spin
interaction, which may compensate largely the antiferromagnetic superexchange
interaction, leading to the predominance of the
biquadratic interaction with $\beta >1$. We believe that this mechanism may be
relevant to LiVGe$_2$O$_6$. In that case the three orbitals come from the $t_{2g}$
orbitals of the vanadium ions, and the degeneracy is lifted due to a small
distortion of the octahedra around the vanadium atoms\cite{millet}. More
generally, we believe that the exchange Hamiltonian is likely to have higher
order contributions whenever the degeneracy between the last occupied and first
empty orbitals is only slightly lifted.  This situation could be realized
in some other transition metal oxides. 

Let us consider a lattice of atoms with two outer electrons per atom.  The
atomic orbitals of the lowest energy level is two-fold degenerate, labelled by
indices 1 and 2.  There is a nearby level of orbital 3, whose energy is higher
by an amount of $\Delta$. We focus on the electron interactions at the same
atom, and denote by $U$ ($U'< U$) the direct Coulomb repulsion between two
electrons in the same (different) orbitals, and by $J_H > 0$ the exchange
interaction. The exchange interaction favors a total spin 1 state for two
electrons on each atom, where orbitals 1
and 2 are both singly occupied with parallel spins, and orbital 3 is empty.
This is in accordance with Hund's rule.  The effective spin-spin coupling
arises when the electron has a virtual transition to the neighboring
atoms. Let $t_{lm}$ be the electron hopping integral from orbital $l$
at site $i$ to orbital $m$ at site $j$ with the electron spin conserved. If the
hopping integral is small compared with $U'$, $J_H$ and $\Delta$, only virtual
transitions are possible, and their net
effect is to induce an intersite spin-spin coupling. 

\begin{figure}[hp]
\centerline{\psfig{figure=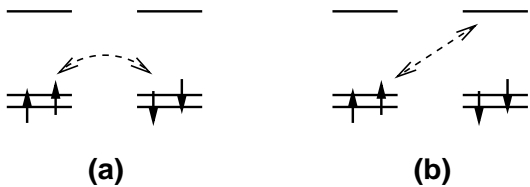,width=7.0cm,angle=0}}
\vspace{0.5cm}
\caption{Typical second--order exchange processes leading to : a) 
Antiferromagnetic bilinear exchange; b) Ferromagnetic bilinear exchange.}
\label{fig2}
\end{figure}

The spin Hamiltonian $H_{ij}$ is uniquely determined by the total spin of 
two neighboring sites $i$ and $j$.  Let $E_S$ the energy of the two-site
system with total spin $S$, with $S$ = 0, 1 or 2. Then  $J_1 = \frac{1}{2}(E_2
-E_1)$,  and
$J_2= \frac{E_0}{3} + \frac{E_2}{6} - \frac{E_1}{2} $. Treating the hopping
integrals as small  parameters, the energy $E_S$ can be calculated within 
perturbation theory.  To illustrate the essential physics, we consider the
simple case where $t_{11}=t_{12}=t_{13}=0$, and $t_{22}=t_{33}$. 
To second order in the hopping integrals, and in units of $t_{22}^2/U'$, 
the spin
couplings are given by~\cite{notes2}
\begin{eqnarray} 
J_1^{(2)} & = & \frac{1}{2}(a_2 +a_3) -\frac{2}{3}\alpha^2 (a_1 -a_2),  
\nonumber \\ 
J_2^{(2)} & = & 0,
\label{J2nd}
\end{eqnarray}
where $\alpha =t_{23}/t_{22}$, and the $a's$ are related to the Hund's
couplings: $a_1 =(1-j_H)^{-1}$, $a_2 =(1+2j_H)^{-1}$,  $a_3=(1+4j_H)^{-1}$,
with $j_H=J_H/U'$.  
From Eqn. (\ref{J2nd}), we see that the biquadratic
interaction vanishes in the second order of perturbation.  This explains the
smallness of the biquadratic coupling in most systems.  The first term in
$J_1^{(2)}$ arises from hopping to an occupied orbital of the neighbouring site
(see Fig. 2a) and corresponds to superexchange interaction. It is positive
and favors 
antiferromagnetic alignment of the two spins. The second term arises from 
the virtual
transition via the orbital state $3$ (see Fig. 2b). Since $a_1 > a_2$, the contribution to
$J_1$ is negative and it favors ferromagnetic alignment of the spins. 
Typical virtual
transitions for the first and the second terms in $J_1$ are illustrated in
Fig. 2.  These two exchange mechanisms may compensate each other, leading to 
a very small
net value of $J_1$. In this situation, it is necessary to extend the
perturbation to include the contributions from the fourth order terms.
Since $\Delta << U', J_H$, the most important contribution arises from 
processes involving a virtual state of spin 1 on both sites and
with one electron in the excited state of orbital $3$. Such a process gives
an energy correction to the second order results by a ratio of order of
$\delta = t^2/U'\Delta$.  In Fig. 3 we show a
typical example of such processes.  All the other fourth order contributions
are smaller by a factor of $\Delta/J_H$ or $\Delta/U'$, and will be neglected 
in the present consideration. 

\begin{figure}[hp]
\centerline{\psfig{figure=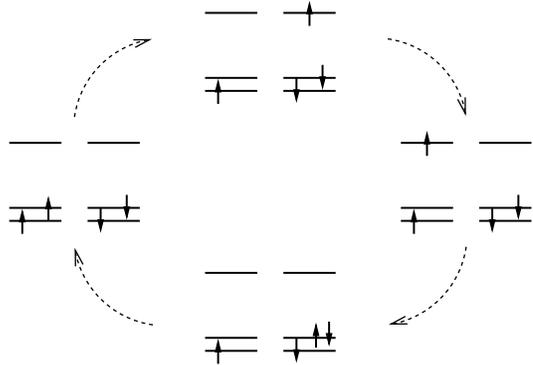,width=7.0cm,angle=0}}
\vspace{0.5cm}
\caption{Typical fourth--order process leading to a biquadratic
interaction.}
\label{fig3}
\end{figure}

Summing over these fourth order terms, we find
\begin{eqnarray}
J_1^{(4)} & = & \delta [ b_2^2 -b_1^2 +(b_3+ \frac{b_1}{3})^2], 
\nonumber \\
J_2^{(4)} & = & - \frac{\delta}{2} [b_2^2 + ( b_3 -\frac{2}{3}b_1)^2],
\label{J4th}
\end{eqnarray}
where $b_1 = \alpha^2 a_1$, $b_2 = \alpha (a_2 +a_3)$, and $b_3=(a_2-a_3)/2
-\alpha^2 a_2/3$. The correction to $J_1$ is simply an energy shift, which
modifies slightly the location of the transition between ferromagnetic and
antiferromagnetic bilinear interaction. 
The biquadratic interaction first appears in the fourth
order in perturbation theory.  While the ratio $\beta$ is usually
small, of the order of $\delta$, it can be significant if the two
contributions in the second order balance out and if $\Delta$ is relatively
small. From equation (\ref{J4th}), we also note that $J_2 < 0$, which 
suggests such a system is
always in the lower half plane of the phase diagram in Fig. 1. This 
also implies
$\beta >0$ if $J_1 >0$, and $\beta <0$ if $J_1 <0$.  In Fig. 4b, we plot the
value of $\beta$ as a function of $\alpha$ for given $\Delta$ and
$J_H/U'$.  At $\alpha < 0.5$, $\beta$ is vanishingly small. As $\alpha$
increases, $\beta$ increases sharply, and the divergent point corresponds to
the transition from antiferromagnetic to ferromagnetic bilinear interaction.

\begin{figure}[hp]
\centerline{\psfig{figure=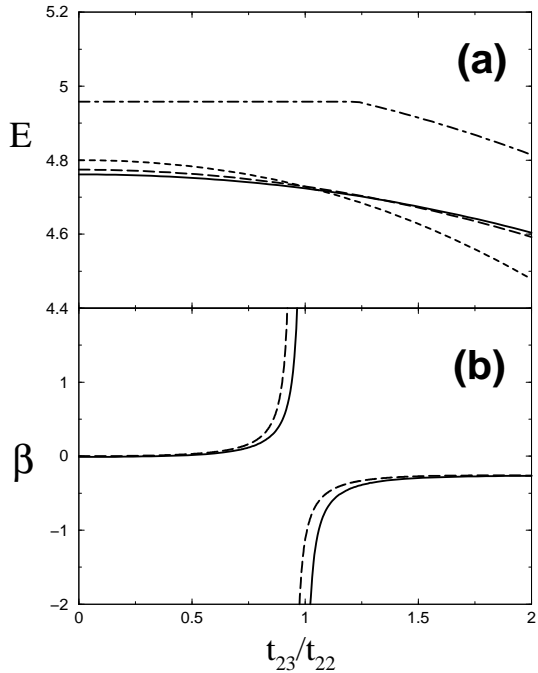,width=7.0cm,angle=0}}
\vspace{0.5cm}
\caption{Exact diagonalization results for a two-site cluster. a) Lowest energy
levels in units of eV. Note the large gap between the three lowest levels
(solid line: S=0; long-dashed line: S=1; short-dashed line: S=2) and the next
one (dashed-dotted line). b)
Relative biquadratic interaction $\beta$ as a function of $t_{23}/t_{22}$ for
typical parameters ($t_{22}=.3$ eV, $\Delta=.3$ eV, $U=6$ eV, $J_H=U/5$). 
Solid line: exact
result; dashed line: perturbation result.}
\label{fig4}
\end{figure}

As stated above, the spin couplings are related to the
energy levels of the two spin 1 system as a function of the total spin. 
These energies can be
numerically calculated exactly. For a two-site system with four electrons and
three orbitals at each site, we have calculated exactly the 495 
energy levels for
various parameters. As
long as $\Delta$ is not too small as compared to the largest hopping integral,
the low-energy part of the spectrum consists of three levels with total spin
$S=0$, $S=1$ and $S=2$ respectively (Fig. 4a). So the low energy effective
Hamiltonian is indeed a spin 1 model.
From the lowest energies $E_S$ for total spin $S=0, 1$ and
$2$, we have extracted $J_1$ and $J_2$. Our numerical results are in good
agreement with the perturbation theory. In particular, we find a transition
from antiferromagnetic to ferromagnetic $J_1$ as a function of
$\alpha=t_{23}/t_{22}$, and a sharp increase in $\beta$ in the same region. 
This transition corresponds to the crossings in Fig. 4a.
The exact numerical results for $\beta$ as function of $\alpha$ for typical
parameters are also plotted in
Fig. 4b. Except for the precise value of $\alpha$ for which $\beta$ diverges, 
the
essential features found in the exact numerical calculations are the same as 
in the perturbation theory. Let us note that the numerical results do not depend
qualitatively on the details of the parameters as long as the fundamental
processes are present: There is always a value of $t_{23}/t_{22}$ around 1 at
which $\beta$ diverges. So although the simplified model presented here in order
to have compact expressions for the fourth order perturbation results
($t_{11}=t_{12}=t_{13}=0$, $t_{33}=t_{22}$) is not exactly realized in
LiVGe$_2$O$_6$, we have checked numerically that the appropriate extension gives
the same behaviour.

The relative value of the biquadratic interaction $\beta$ depends 
sensitively on the ratio of the
hopping integrals, but not on their magnitude.
Application of a pressure on the system, if it is uniformly added, 
may not change the ratio of these hopping integrals, hence the biquadratic
interaction sensitively. A uniaxial pressure, on the other hand, may change the
hopping integrals in different proportions, and could dramatically affect the
magnetic  properties of the system. This possibility would be worth
investigating in the case of LiVGe$_2$O$_6$.

Finally, let us put these results in perspective. Although this mechanism
would apply in any dimension, there is something very special about 1D.
In higher dimension, fourth--order perturbation theory produces other types of
four-spin interactions involving spins at four different sites. However, in 1D 
these processes do not appear as long as the hopping to further neighbours can
be neglected, which usually is the case. So this mechanism is somehow specific
to 1D. This might be the reason why it was overlooked in early attempts at
producing strong biquadratic interactions because they were done before
Haldane's conjecture and were more concerned with 3D magnets. 

This mechanism is also somehow related to the problem of orbital degeneracy. It
is well known that orbital degeneracy leads to a Hamiltonian which is not purely
Heisenberg, but involves a pseudo-spin for the orbital degrees of
freedom\cite{kugel}. For spin 1/2 systems, for which most of the work has been
done, the bilinear interaction is the only candidate - higher order terms do not
appear because they can be rewritten in terms of the bilinear interaction - and 
the situation is a clear cut: Either the orbitals do not play a role, and the
Hamiltonian is purely Heisenberg, or they do, and the Hamiltonian is a
spin--orbital model. For spin 1 systems, we showed that, on going from a
Heisenberg to a spin--orbital model by pulling down an empty orbital, 
there is an intermediate region where the Hamiltonian is still a pure spin
Hamiltonian, but with a more general interaction than just bilinear due to a
non-trivial level crossing (Fig. 4a). This simple
idea should be useful in other contexts as well in looking for general spin
Hamiltonians. In that respect, we hope that the present results will encourage
quantum chemists to look more closely at exchange in situations where orbitals
are quasi-degenerate. Quantum chemistry has proven to be able to reproduce
very accurately on the basis of
{\it ab initio} calculations the values of the exchange integrals 
of several
systems. It should thus be possible along the same lines to specify more
precisely the conditions under which the mechanism proposed in this paper will
apply. 

In conclusion, we have provided a simple but efficient mechanism to produce
significant biquadratic interactions in spin 1 chains. It is our hope 
that these ideas will ultimately lead to an experimental investigation of the
fascinating phase diagram of the general spin 1 model in one dimension.

NOTE: After completion of this project, we became aware of a paper by Bhatt and
Yang [R. N. Bhatt and K. Yang, J. Appl. Phys. {\bf 83}, 7231 (1998)] who
considered a similar problem in the context of random antiferromagnetic spin 
chains. Their model is slightly different from ours (they consider the case of 
$N$ degenerate orbitals with a simplified interaction), and they restrict
themselves to a perturbation calculation, but our results are
qualitatively consistent with theirs.

ACKNOWLEDGEMENTS: We acknowledge useful discussions with T. Jolicoeur, K. Le Hur
and O. Golinelli. This work was supported in part by DOE Grant No.
DE/FG03-98ER45687.


\begin{thebibliography}{4}
\bibitem{EuSe} One of the exceptions is the 3-dimensional compound $EuSe$, 
but with a mechanism different from the one discussed in this paper.
\bibitem{notes2} Here we consider the case $U=U'+2J_H$, and  $\Delta << U',J_H$.
\end{thebibliography}
\end{document}